\documentclass[a4paper, oneside, 11pt]{article}
\usepackage[english]{babel}
\usepackage{amsfonts}
\usepackage{mathrsfs}
\usepackage{amssymb}
\usepackage{graphics}
\usepackage{amsrefs}
\usepackage{latexsym}
\begin{document}

\title{{\bf{\Large{Yet another application of marginals of multivariate Gibbs distributions}}}\footnote{{\it AMS (2000) subject classification}. Primary: 60G58. Secondary: 60G09.}}
\author{\textsc {Annalisa Cerquetti}\footnote{Corresponding author, SAPIENZA University of Rome, Via del Castro Laurenziano, 9, 00161 Rome, Italy. E-mail: {\tt annalisa.cerquetti@gmail.com}}\\
\it{\small Department of Methods and Models for Economics, Territory and Finance}\\
  \it{\small Sapienza University of Rome, Italy }}
\newtheorem{teo}{Theorem}
\date{20 Dicembre 2013}
\maketitle{}

\begin{abstract}
We give yet another example of the usefulness of working with {\it marginals of multivariate Gibbs distributions} (Cerquetti, 2013) in deriving Bayesian nonparametric estimators under Gibbs priors in species sampling problems. Here in particular we substantially reduce length and complexity of the proofs in Bacallado {\it et al.} (2013, Th. 1, and Th. 2) for looking backward probabilities under incomplete information. 
\end{abstract}

\section{Introduction} In the typical setting of Bayesian nonparametric inference  for species sampling problems  under Gibbs priors (Lijoi {\it et al.} 2007, 2008, Favaro {\it et al.} 2012, 2013) 
Bacallado {\it et al.}  (2013) obtain conditional falling factorial moments, given an initial sample of observations, for  the number of old species re-observed and for the number of old species re-observed a certain number of times in an additional sample,  under {\it complete and incomplete} information. As the Authors explain, here {\it incomplete information} stands for the possibility to know  the number $j$ of different species observed in the initial $n$-sample, but not the specific multiplicities $(n_1, \dots, n_j)$. 

Recently it has been shown that relying on marginals of multivariate Gibbs distributions, posterior predictive inference in species sampling problems can be considerably simplified (Cerquetti, 2013, 2013b). Here we give another example of the usefulness of this new technique   deriving the main results in Bacallado {\it et al.} (2013) quickly and easily.  
For the sake of brevity we just recall here the essential preliminaries and notation.  We refer the interested reader to the cited literature for a comprehensive introduction to BNP in species sampling.

\section {Some preliminaries on notation}

Gibbs priors are a large class of laws for infinite discrete distributions $(P_i)_{i \geq 1}$ introduced by Gnedin and Pitman (2006) as generalizations of the two-parameter Poisson-Dirichlet priors (Pitman and Yor, 1997). The Gibbs class is conveniently identified by the corresponding  family of exchangeable partition probability functions (EPPFs) it induces by random sampling. Those are distributions on the space of consistent sequences of random {\it partitions} $\{A_1, \dots, A_j\}$ of the finite sets $[n]:=\{1, \dots, n\}$  
characterized by the Gibbs product form 
$$
p_{\alpha, V}(n_1, \dots, n_j)=V_{n,j}\prod_{i=1}^j (1-\alpha)_{n_i-1},
$$ 
where $p$ is a symmetric function of {\it compositions} $(n_1, \dots, n_j)$ of $n$ corresponding to the sizes of the $j$ different blocks  in order of appearance,  $V=(V_{n,j})$ are weights that identify the specific Gibbs model and $\alpha \in (-\infty, 1)$.  Notice that here, with respect to Bacallado {\it et al.} (2013) we use $\alpha$ in place of $\sigma$ as it is standard in exchangeable random partitions literature and  the lighter notation $(x)_s=(x)(x+1) \cdots(x +s -1)$ for rising factorials. By Theorem 12 in Gnedin and Pitman (2006) each exchangeable Gibbs partition arises as a probability mixture of extreme partitions, which differ for $\alpha <0$, $\alpha=0$ and $\alpha \in (0,1)$. The joint law of the corresponding random vector $(N_1, \dots, N_{K_n}, K_n)$ of the blocks' multiplicities in exchangeable random order and the number of blocks, first introduced in Pitman (2006, eq. (2.7)),  has been termed {\it multivariate Gibbs distribution} in Cerquetti (2013) and is given by 
\begin{equation}
\label{multiG}
\mathbb{P}_{\alpha, V} (N_1=n_1, \dots, N_{K_n}=n_j, K_n=j)= \frac{n!}{\prod_i n_i! j!}V_{n,j} \prod_{i=1}^j (1-\alpha)_{n_i-1}.
\end{equation}
The law of $K_n$ follows  by marginalizing (\ref{multiG}) over the space of {\it compositions} of $n$ with $j$ blocks 
\begin{equation}
\label{leggekappa}
\mathbb{P}_{\alpha, V}(K_n=j)= V_{n, j} S_{n, j}^{-1, -\alpha}
\end{equation}
where $S_{n,j}^{-1, -\alpha}$ are generalized Stirling numbers
as arising from Bell polynomials
\begin{equation}
\label{bell_uno}
B_{n, j}(w_{\bullet})=\frac{n!}{j!} \sum_{(n_1, \dots, n_j)} \prod_{i=1}^j \frac{w_{n_i}}{n_i!}
\end{equation}
for $w_{\bullet}=(1-\alpha)_{\bullet-1}$.\\\\
Notice that in Bacallado {\it et al.} (2013) equation (2) is expressed in terms of {generalized factorial coefficients} $\mathcal{C}(n, j; \alpha)$ as
\begin{equation}
\label{leggekappaLORO}
\mathbb{P}_{\alpha, V}(K_n=j)= V_{n, j} \frac{\mathcal{C}(n, j; \alpha)}{\alpha^j}
\end{equation}
where $\mathcal{C}(n,j; \alpha)$ are the connection coefficients
\begin{equation}
\label{genfac}
(\alpha x )_n= \sum_{j=0}^n \mathcal{C}(n, j; \alpha) (x)_j,
\end{equation}
and are not defined for $\alpha=0$. Since the general Gibbs class is defined for $\alpha \in (-\infty, 1)$ here, as in Cerquetti (2013, 2013b) we use generalized Stirling numbers  $S_{n,j}^{-1, -\alpha}$ defined as connection coefficients 
$$
(x)_n=\sum_{j=0}^n S_{n,j}^{-1, -\alpha} (x)_{j \uparrow \alpha}
$$
where $(x)_{j \uparrow \alpha}=x(x+\alpha)(x +2\alpha)\cdots(x +(j-1)\alpha)$ are generalized rising factorials.  For $\alpha \neq 0$ then $S_{n,j}^{-1, -\alpha}= \frac{\mathcal{C}(n,j; \alpha)}{\alpha^j}$. For $\alpha=0$ then $(x)_{j\uparrow 0}=x^j$ and those numbers correspond to  {\it signless Stirling numbers} of the first kind $S_{n,j}^{-1, 0}$.

\section{Looking backward via marginals of multivariate Gibbs}
In species sampling problems given an initial sample of size $n$ with $j$ different species observed with multiplicities $(n_1, \dots, n_j)$, interest usually lies on inferring the behaviour of an additional sample of size $m$ with respect to the number $K_m$ of {\it new} species, the total number $L_{m}$ of observations belonging to new species,  the multiplicities $M_1, \dots, M_{K_m}$ of the new species observed, and  the number $S_1, \dots, S_{j}$ of new observations belonging to {\it old species}.  (See Cerquetti, 2013, Lijoi {\it et al.}, 2007, 2008, Favaro {\it et al.}, 2012, 203). 
The main results in Bacallado {\it et al.} (2013) are in  Theorem 1. and Theorem 2. which provide respectively:
\begin{itemize}
\item conditional falling factorial moments for $R_{m}^{(j,n, {\bf n})}$ and $R_{m}^{(j,n)}$,  the total number of old species re-observed in the additional sample under {\it complete} and {\it incomplete} information (Theorem 1, eq. (3.4) and (3.5)).
\item conditional falling factorial moments for $R_{l, m}^{(j,n, {\bf n})}$ and $R_{l, m}^{(j,n)}$ the number of old species re-observed $l$ times in the additional sample under {\it complete} and {\it incomplete} information (Theorem 2, eq. (3.12) and (3.13)).
\end{itemize}

To obtain results (3.4) and (3.12), as the same Authors state in the proofs, it is enough to specialize for $l=0$ the result in Theorem 1 in Favaro {\it et al.} (2013) and to resort to $
R_{m}^{(j,n, {\bf n})}= j - R_{0,m}^{(j,n,{\bf n} )}
$. See  also Cerquetti (2013, eq. 26) for a result in terms of generalized Stirling numbers.  

To obtain results (3.5) and (3.13) the Authors adopt a complicated procedure resulting in two different proofs of about six pages each. Here we present a far more easy route relying on marginals of multivariate Gibbs distributions as introduced in Cerquetti (2013). Previous examples of the usefulness of this new technique are in Cerquetti (2013b).

First notice that since the total number of old species is observed, then even under incomplete information 
$$
R_{m}^{(j,n)}= j - R_{0,m}^{(j,n)}.
$$
Therefore, by properties of rising factorials and the definition of {\it non central Lah numbers}, which correspond to generalized Stirling numbers for $\alpha=-1$,
\begin{equation}
\label{lah_2}
S_{r,v}^{-1, 1}= \frac{r!}{v!}{r-j-1 \choose r-v}
\end{equation}
then
\begin{equation}
\label{lah_1}
[(R_{m}^{(j,n)})_{r \downarrow }]= [(j - R_{0,m}^{j,n})_{r \downarrow }]=(-1)^r\sum_{v=0}^r S_{r,v}^{-1,1, j}(R_{0,m}^{j,n})_{v \downarrow }.
\end{equation}
This implies that it is enough to give a proof for (3.13), since  (3.5) follows easily.  \\\\
{\bf Remark 1.} Eq. (3.13) in Bacallado {\it et al.} (2013) is written in terms of {\it generalized factorial coefficients} $\mathcal{C}(n, j; \alpha)$ (central and non central) namely
\begin{equation}
\label{laloro313}
\mathbb{E}[(R_{l,m}^{(n,j)})_{r \downarrow}]
=\frac{r!}{\mathcal{C}(n,j, \alpha)}{\frac{m!}{(l!)^r (m-rl)!}} [-\alpha(1-\alpha)_{l-1}]^r $$
$$
\times \sum_{s=r}^{n-(j-r)}{n \choose s} \mathcal{C}(s,r, \alpha-l) \mathcal{C}(n-s, j-r, \alpha)
\end{equation}
$$
\times \sum_{k=0}^{m-rl} \frac{V_{n+m, j+k}}{V_{n,j}} \frac{\mathcal{C}(m-rl, k, \alpha, -(n-s-(j-r)\alpha)}{\alpha^k}.
$$
Since generalized factorial coefficients  are not defined for $\alpha=0$   the previous expression is not directly applicable for posterior inference under priors belonging to the Gibbs class for  $\alpha=0$. To recover the value of the estimator for $\alpha=0$ it is necessary to resort to the limit value for $\alpha \rightarrow 0$ exploiting the known relationship 
$$
\lim_{\alpha \rightarrow 0}\frac{\mathcal{C}(n,j, \alpha)}{\alpha^j}=S_{n,j}^{-1, 0},
$$
where $S_{n,j}^{-1,0}$ are {\it signless Stirling numbers of the first kind}.
The very same problem applies to equations (3.4), (3.5) and (3.12).  To avoid this kind of drawbacks we advocate here the use of generalized Stirling numbers in Bayesian nonparametrics under Gibbs priors, as it is standard in the exchangeable Gibbs partitions literature (cf. e.g. Gnedin and Pitman, 2006; Pitman, 2006). We stress  that, despite still not completely explored, the class of Gibbs priors with $\alpha=0$ is theoretically infinite. The corresponding Gibbs weights $V_{n,k}$ can be obtained mixing  over $\theta$ the weights of Dirichlet priors $V_{n,k}^{\theta}=\frac{\theta^k}{(\theta)_n}$ with a general density  $\gamma(\theta)$ on $(0, \infty)$. \\\\

The following Proposition gives an easy and short proof for $\mathbb{E}[(R_m^{(j,n)})_{r \downarrow}]$ as obtained in eq. (3.13) in Bacallado {\it et al.} (2013).\\\\
{\bf Proposition 1.} {\it Let $n$ and $j$ be respectively the size and the number of different species observed in a sample from an unknown population of infinite species. Let $m$ be the size of an additional sample, then the $r$-th falling factorial moment of the number $R_{l,m}$ of old species reobserved $l$ times, given  $n$ and $j$ under  $(\alpha, V)$-Gibbs priors, for $\alpha \in (-\infty, 1)$, is given by}
\begin{equation}
\label{laloro313mia}
\mathbb{E}^{\alpha, V}[(R_{l,m}^{(n,j)})_{r \downarrow}]
={r!}{\frac{m!}{(l!)^r (m-rl)!}} [(1-\alpha)_l]^r 
\times \sum_{s=r}^{n-(j-r)}{n \choose s} \frac{S_{s,r}^{-1, -(l-\alpha)} S_{n-s, j-r}^{-1, -\alpha}}{S_{n,j}^{-1, -\alpha}}
\end{equation}
$$
\times\sum_{k=0}^{m-rl} \frac{V_{n+m, j+k}}{V_{n,j}} {S_{m-rl, k}^{-1, -\alpha, -(n-s-(j-r)\alpha)}}.
$$
{\it Proof.} Let ${\bf N}=(N_1, \dots, N_j, K_n=j)$ then 
by definition of {\it incomplete information}
$$
\mathbb{E}^{\alpha, V}[(R_{(l, m)}^{n,j})_{r \downarrow }]= \mathbb{E}^\alpha_{{\bf N}|K_n=j} \left[\mathbb{E}^{\alpha, V}[(R_{l, m}^{n,j, {\bf N}})_{r \downarrow }]\right]
$$
where, by (3.12) in Bacallado {\it et al.} (2013) expressed in terms of generalized Stirling numbers 
$$
 \mathbb{E}^{\alpha, V}[(R_{l,m}^{(n,j, {\bf N})})_{r \downarrow}]= r! \frac{m!}{(l!)^r (m-rl)!} \sum_{(c_1, \dots, c_r) \in C_{j,r}} \prod_{i=1}^r (N_{c_i}-\alpha)_{l}$$
 $$
 \times
 \sum_{k=0}^{m-rl} \frac{V_{n+m, j+k}}{V_{n,j}}S_{m-rl, k}^{-1, -\alpha, -(n-\sum_i N_{c_i}-(j-r)\alpha)}.
$$
Now for $T_r=\sum_{i=1}^r N_{c_i}$ taking values in $[r, n-(j-r)]$, we can write in short form
$$
 \mathbb{E}^{\alpha, V}[(R_{l,m}^{(n,j, {\bf N})})_{r \downarrow}]= \frac{r!m!}{(l!)^r (m-rl)!}\sum_{(c_1, \dots, c_r) \in C_{j,r}} \left[\prod_{i=1}^r (N_{c_i}-\alpha)_l\right]g_{m,r,l,j}^{\alpha, V}(T_r) 
$$
where $C_{j,r}$ is the space of $r$-combinations of $[j]$, hence
$$
 \mathbb{E}^\alpha_{{\bf N}|K_n=j} \left[\mathbb{E}[(R_{l, m}^{(n,j, {\bf N})})_{r \downarrow }]\right]= \mathbb{E}^{\alpha}_{{N_{c_1}, \dots, N_{c_r}|K_n=j}} \left[\mathbb{E}^{\alpha, V}[(R_{l, m}^{(n,j, {\bf N})})_{r \downarrow }]\right]=
$$
and by exchangeability
\begin{equation}
\label{prima}
= {j \choose r}\frac{r!m!}{(l!)^r (m-rl)!}\mathbb{E}^{\alpha}_{{N_{1}, \dots, N_{r}|K_n=j}}\left\{\left[\prod_{i=1}^r (N_{i}-\alpha)_l\right] g_{m,r,l,j}^{\alpha, V}(T_r) \right\}.
\end{equation}
By (7) in Cerquetti (2013) and (\ref{leggekappa}) for $r \leq j$
$$
\mathbb{P}_{\alpha}(N_1=n_1, \dots, N_r=n_r|K_n=j)= \frac{n!}{\prod_{i=1}^r n_i! (n-\sum_i n_i)! } \frac{\prod_{i=1}^r (1-\alpha)_{n_i-1}}{j_{[r]}} \frac{S^{-1, -\alpha}_{n -\sum_i n_i, j-r}}{S_{n,j}^{-1, -\alpha}}.
$$
Now notice that $(n_i -\alpha)_{l}(1-\alpha)_{n_i-1}= (1-\alpha)_{l}(1-\alpha+l)_{n_i-1}$, hence (\ref{prima}) corresponds to
$$= {j \choose r}\frac{r!m!}{(l!)^r (m-rl)!}[(1-\alpha)_l)^r]\mathbb{E}^{\alpha-l}_{{N_{1}, \dots, N_{r}|K_n=j}} [g_{m,r,l,j}^{\alpha, V}(T_r)]  \frac{S^{-1, -(\alpha-l)}_{n, j}}{S_{n-\sum_{n_i},j-r}^{-1, -(\alpha-l)}}\frac{S^{-1, -\alpha}_{n -\sum_i n_i, j-r}}{S_{n,j}^{-1, -\alpha}} $$
and
\begin{equation}
\label{seconda}
={j \choose r}\frac{r!m!}{(l!)^r (m-rl)!}[(1-\alpha)_l)^r]\mathbb{E}^{\alpha-l}_{T_r|K_n=j} [g_{m,r,l,j}^{\alpha, V}(T_r)] \frac{S^{-1, -(\alpha-l)}_{n, j}}{S_{n-\sum_{n_i},j-r}^{-1, -(\alpha-l)}}\frac{S^{-1, -\alpha}_{n -\sum_i n_i, j-r}}{S_{n,j}^{-1, -\alpha}}.
\end{equation}
By equation (9) in Cerquetti (2013) and the definition of generalized Stirling numbers in terms of Bell polynomials (\ref{bell_uno}), multiplying and dividing by $s!$ and $r!$
$$
\mathbb{P}_{\alpha -l}(T_r=s|K_n=j)=\sum_{\{(n_1, \dots, n_r): \sum_{i} n_i=s\}} \mathbb{P}_{\alpha - l}(N_1, \dots, N_r|K_n=j)= 
$$
$$
= \sum_{\{(n_1, \dots, n_r): \sum_{i} n_i=s\}} \frac{ n! (j-r)! \prod_{i=1}^r(1-\alpha+l)_{n_i -1}S_{n-s, j-r}^{-1, -\alpha+l}}{\prod_{i=1}^{r} n_i! (n-s)! j! S_{n,j}^{-1, -\alpha+l}}={n \choose s}{j \choose r}^{-1}\frac{S_{s,r}^{-1, -\alpha+l}S_{n-s, j-r}^{-1, -\alpha+l}}{S_{n, j}^{-1, -\alpha+l}}.
$$
Therefore, writing explicitly (\ref{seconda}) yields
$$
=\frac{r!m!}{(l!)^r (m-rl)!}[(1-\alpha)_l)^r]\sum_{s=r}^{n-(j-r)}{n \choose s}\frac{S_{s,r}^{-1, -\alpha+l}S_{n-s, j-r}^{-1, -\alpha}}{S_{n,j}^{-1, -\alpha}}[g_{m,r,l,j}^{\alpha, V}(s)] .
$$
{\bf Corollary 1.} For $l=0$  (\ref{laloro313mia}) yields 
$$
[(R_{0,m}^{(j,n)})_{r \downarrow }]=r!\sum_{s=r}^{n-(j-r)}{n \choose s}\frac{S_{s,r}^{-1, -\alpha+l}S_{n-s, j-r}^{-1, -\alpha}}{S_{n,j}^{-1, -\alpha}}g_{m,r,l,j}^{\alpha, V}(s) .
$$
Applying (\ref{lah_1})
$$
[(R_{m}^{(j,n)})_{r \downarrow }]= [(j - R_{0,m}^{(j,n)})_{r \downarrow }]=(-1)^r\sum_{v=0}^r S_{r,v}^{-1,1, j}[(R_{0,m}^{(j,n)})_{v \downarrow } ]
$$
and by (\ref{lah_2}) and recalling that $(x)_s=(-1)^s(-x)_{s \downarrow}$
equation (3.5) in Theorem 1. in Bacallado {\it et al.} (2013) rewritten in terms of generalized Stirling numbers is recovered by elementary combinatorics.

\section*{Acknowledgements} The author wishes to thank Lorenzo Trippa for kindly explaining the functioning of formulas (3.4), (3.5), (3.12) and (3.13) in Bacallado {\it et al.} (2013) under $\alpha=0$.

\section*{References}
\newcommand{\bibu}{\item \hskip-1.0cm}
\begin{list}{\ }{\setlength\leftmargin{1.0cm}}

\bibu \textsc{Bacallado, S., Favaro S. and Trippa, L.} (2013) Looking-backward probabilities for Gibbs type exchangeable random partitions. {\it Bernoulli} (to appear).

\bibu \textsc{Cerquetti, A.} (2013) Marginals of multivariate Gibbs distributions with applications in Bayesian species sampling {\it Elect. J. Stat.},7,  697--716.

\bibu \textsc{Cerquetti, A.} (2013b) A note on a Bayesian nonparametric estimator of the discovery probability. {\it arXiv:1304.1030} [math.ST]

\bibu \textsc{Favaro, S., Lijoi, A. and Pr\"unster, I.} (2012) A new estimator of the discovery probability. {\it Biometrics, {\bf 68}, 1188-1196}.

\bibu \textsc{Favaro, S., Lijoi, A. and Pr\"unster, I.} (2013) Conditional formulae for Gibbs-type exchangeable random partitions. {\it Ann. Appl. Probab.} {\bf 23}, 5, 1721--1754.

\bibu \textsc{Gnedin, A. and Pitman, J. } (2006) {Exchangeable Gibbs partitions  and Stirling triangles.} {\it Journal of Mathematical Sciences}, 138, 3, 5674--5685. 

\bibu \textsc{Lijoi, A., Mena, R.H. and Pr\"unster, I.} (2007) Bayesian nonparametric estimation of the probability of discovering new species.  {\it Biometrika}, 94, 769--786.

\bibu \textsc{Lijoi, A., Pr\"unster, I. and Walker, S.G.} (2008) Bayesian nonparametric estimator derived from conditional Gibbs structures. {\it Annals of Applied Probability}, 18, 1519--1547.


\bibu \textsc{Pitman, J.} (2006) {\it Combinatorial Stochastic Processes}. Ecole d'Et\'e de Probabilit\'e de Saint-Flour XXXII - 2002. Lecture Notes in Mathematics N. 1875, Springer.

\bibu \textsc{Pitman, J. and Yor, M.} (1997) The two-parameter Poisson-Dirichlet distribution derived from a stable subordinator. {\it Ann. Probab.}, 25, 855--900.

\end{list}

\end{document}